\documentclass[twocolumn,
aps,
prb,
showpacs,
amsmath,
amssymb,
floatfix,
raggedbottom,
nobalancelastpage,
reprint,
superscriptaddress,
eprint]{revtex4-1}

\pdfoutput=1

\usepackage{xcolor,mathrsfs,dsfont}
\definecolor{darkblue}{RGB}{0,0,150}
\definecolor{nightblue}{RGB}{0,0,100}

\usepackage{graphicx,mathtools,bm}
\usepackage[
colorlinks,
citecolor=darkblue,
linkcolor=darkblue,
urlcolor=nightblue]{hyperref}

\usepackage[english]{babel}
\usepackage[babel,kerning=true,spacing=true]{microtype}

\usepackage{appendix}
\usepackage{subfigure}

\newcommand{\dd}{\mathrm{d}}
\renewcommand{\Im}{\ensuremath{\mathrm{Im}\;}}
\renewcommand{\Re}{\ensuremath{\mathrm{Re}\;}}

\bibpunct{[}{]}{,}{n}{}{}
\makeatletter
\def\NAT@def@citea{\def\@citea{\NAT@separator}}
\makeatother

\begin{document}

\title{Hydrodynamic Coulomb drag
and bounds on diffusion}

\author{Tobias Holder}
\affiliation{Department of Condensed Matter Physics, Weizmann Institute of Science, Rehovot, Israel 7610}
\date{\today}

\begin{abstract}
We study Coulomb drag between an active layer with a clean electron liquid and a passive layer with a pinned electron lattice in the regime of fast intralayer equilibration. 
Such a two-fluid system offers an experimentally realizable way to disentangle the fast rate of intralayer electron-electron interactions from the much slower rate of momentum transfer  between both layers.
We identify an intermediate temperature range above the Fermi energy of the electron fluid but below the Debye energy of the electronic crystal
where the hydrodynamic drag resistivity is directly proportional to a fast electron-electron scattering rate.
The results are compatible with the conjectured scenario for strong electron-electron interactions which poses that a linear temperature dependence of resistivity originates from a ``Planckian'' electron relaxation time $\tau_{eq}\sim \hbar/k_BT$. 
We compare this to the better known semiclassical case, where the diffusion constant is found to be not proportional to the microscopic timescale. 
\end{abstract}

\maketitle

\section{Introduction}
Hydrodynamic transport in interaction dominated electron liquids has emerged as a tantalizing testing ground for non-Fermi liquid phenomena. 
To characterize this transport regime, a conjecture has attracted attention which relates the diffusion constant to a yet to be determined microscopic velocity $v$ and equilibration time $\tau_{eq}$,
\begin{align}
D\sim v^2\tau_{eq}.
\label{eq:simplediffusion}
\end{align}
While such a relation follows trivially from a Drude-type model for diffusive transport,
a version for interaction dominated systems was first formulated for certain holographic approaches and in a number of strongly correlated model systems~\cite{Son2007,Hartnoll2018}. In these instances, Eq.~(\ref{eq:simplediffusion}) presents a lower bound for the momentum diffusivity.
More recently, the same kind of phenomenology was put forward as a universal signature of hydrodynamic transport in systems where the momentum of a quasiparticle is much longer lived than its energy~\cite{Hartman2017}. In this context, Eq.~(\ref{eq:simplediffusion}) has been considered as a upper bound on momentum diffusivity. While both cases involve very different microscopic mechanisms, they have in common that the microscopic quantities $v$ and $\tau_{eq}$ remain hard to pin down theoretically.

A related conjecture for the electrical resistivity in strongly interacting systems reads~\cite{Hartnoll2015}
\begin{align}
\rho&=\frac{1}{\chi}\frac{1}{v^2}\frac{1}{\tau_{eq}}\sim T,
\label{eq:planckformula}
\end{align}
where the left equality defines the resistivity $\rho$ by use of the charge susceptibility $\chi$. At low temperatures, $\chi$ and $v$ are independent of temperature. The linear-T dependence of the resistivity, observed in many strange metals~\cite{Gunnarsson2003,Bruin2013} is therefore frequently associated with an equilibration time of the order~\cite{Damle1997,Zaanen2004}
\begin{align}
\tau_{eq}\sim\frac{\hbar}{k_B T}.
\end{align}

Both conjectures for diffusivity and resistivity are expected to hold in an intermediate temperature regime, where electron-electron interactions prevail over other relaxation channels. At lower temperatures, they do not hold because the electron-electron scattering rate $\tau_{ee}^{-1}$ decreases and eventually becomes smaller than the disorder scattering rate $\tau_{dis}^{-1}$. Similarly, at high enough temperatures other relaxation channels like umklapp scattering or electron-phonon interactions become important. It is not known whether either relation, Eq.~(\ref{eq:simplediffusion},\ref{eq:planckformula}) can be derived given the other~\cite{Lucas2017a}.

Microscopically, the resistivity as suggested by Eq.~(\ref{eq:planckformula}) was successfully constructed for a Fermi liquid with long wavelength spatial inhomogeneities~\cite{Lucas2018}, which fulfills the criteria for hydrodynamic transport. It is also known that some multi-band models which have both heavy and light bands yield $\rho\sim\tau_{eq}^{-1}$ even in the Fermi liquid regime of weak interactions~\cite{Pal2012}. As an important counterexample, hydrodynamic Coulomb drag between two layers of a quantum critical metal was shown to produce a dissimilar temperature dependence~\cite{Patel2017}.

A number of models are known to give rise to the phenomenology of Eq.~(\ref{eq:planckformula}) by strongly coupling fermions to phonons~\cite{Werman2016,Werman2017}, or to SYK-fermions~\cite{Song2017,Chowdhury2018}.
In these cases momentum dissipation happens through strong umklapp scattering at rates comparable to energy relaxation so that transport is not hydrodynamic. 
However, these models have exposed a number of candidates for the microscopic content of the velocity $v$ in Eq.~(\ref{eq:simplediffusion}). 
Among them are the Fermi velocity $v_F$, sound velocity $v_s$, operator growth velocity $v_{LC}$ or butterfly velocity $v_B$. More
recently it was argued that it is the largest of these velocities which sets the timescale~\cite{Lucas2017a,Lucas2017}.

\begin{figure}
\includegraphics[width=.75\columnwidth]{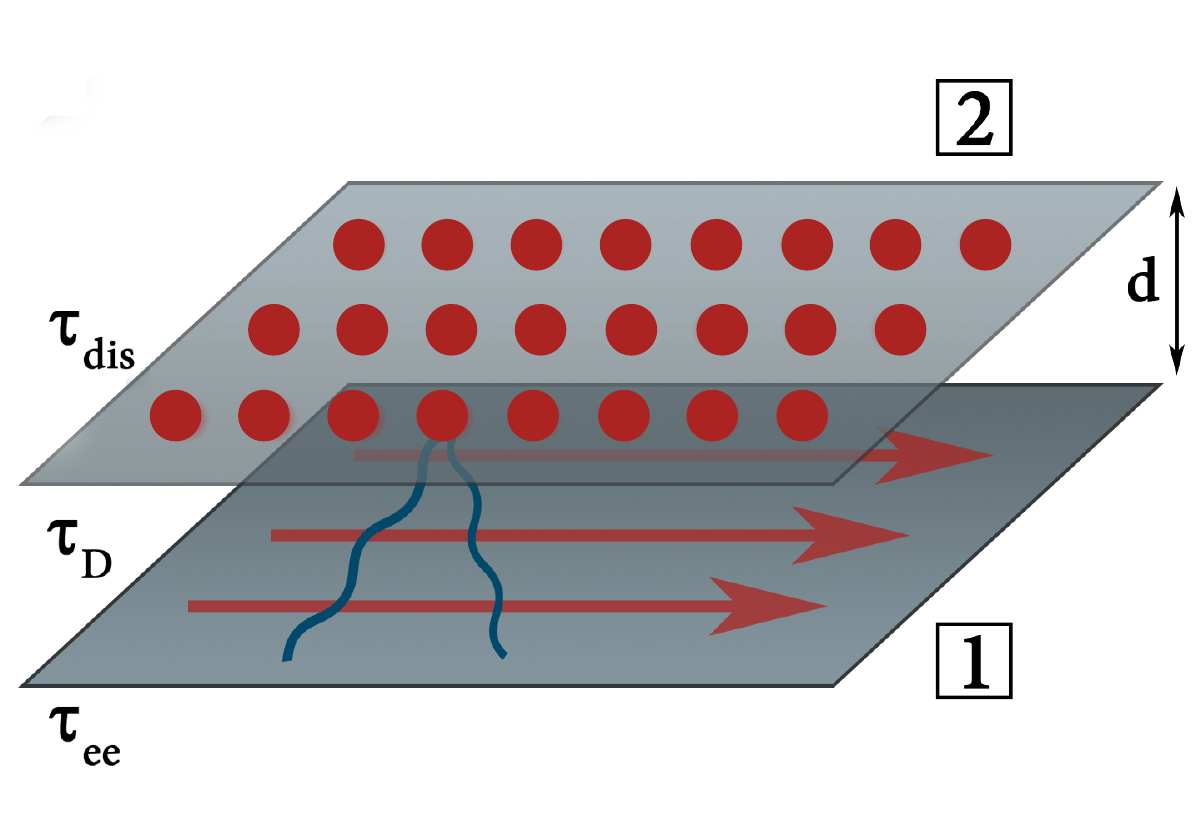}
\caption{Bilayer setup with unequal layers. In the lower layer interacting electrons with lifetime $\tau_{ee}$ carry a current. In the upper layer, separated by a distance $d$, an electronic lattice acts as a phonon bath which dissipates momentum within $\tau_{dis}$. Between both layers, momentum and energy is exchanged by Coulomb interaction with a timescale $\tau_{D}$.}
\end{figure}
Despite these efforts, it is still largely unknown whether a simple one-band Fermi liquid can reach a universal regime of transport, and which microscopic quantities constitute $v$ and $\tau_{eq}$ in this case. At the heart of the problem lies the fact that a substantial electron-electron interaction simultaneously decreases quasiparticle lifetimes and increases screening length, thus producing complicated retardation effects. The problem is essentially intractable unless these effects can be disentangled.

Here, we suggest a Coulomb drag experiment using a Wigner crystal to achieve such a scale separation, which allows for a tunable control of the momentum relaxation rate compared to the electron-electron scattering rate for strongly interacting, clean Coulomb liquids. 
We address in particular how energy and momentum relaxation are related in this setup in the absence of fast extrinsic relaxation mechanisms, where the electron liquid is instead thermodynamically coupled to a bath. 
On the technical side, we put forth a self-consistent formalism utilizing recent developments regarding hydrodynamic drag in the framework of a Boltzmann-Langevin kinetic equation~\cite{Apostolov2014,Chen2015}. 
While this framework remains perturbative in terms of the absolute interaction parameter $r_s$, it can be employed to explore the essentially unscreened limit at low electron density. 
The results presented here show that the parameter regime where temperature, Fermi energy and interaction strength are comparable is not entirely inaccessible, as it was previously thought~\cite{Spivak2006,Andreev2011}.

The main ingredient is to spatially separate the current-carrying electron fluid from its source of dissipation by a distance $d$, thereby introducing a scale separation between 
momentum conserving electron-electron scattering and dissipative scattering, which however both arise from the same Coulomb interaction. 
The active layer (index $1$) is a clean metal with low carrier density which carries a current $\bm{j}_1$. The passive layer (index $2$) is a pinned Wigner crystal in which momentum transport by fermionic quasiparticles is prevented. Importantly, by using electronic lattice excitations as a phonon bath, we evade the complicated electron-core interactions which normally apply for atomic acoustic phonons~\cite{Ziman1960}. 

The proposed system allows to investigate an aspect of Eq.~(\ref{eq:planckformula}) which was only little discussed so far~\cite{Mousatov2018}. If the electron liquid is heated above the Fermi energy ($T>E_F$), the occupied density of states $\chi(T)$ decreases with $T^{-1}$ while the thermal velocity $v(T)$ increases like $T^{1/2}$. The product $\chi(T)v^2(T)$ thus remains constant. At low densities, where the bare interaction scale is substantially larger than $E_F$, it is therefore reasonable not only to expect that Eq.~(\ref{eq:planckformula}) holds also at $T>E_F$ but more importantly that the resistivity still shares the temperature dependence of the microscopic scattering rate. 

We test this hypothesis in the Coulomb drag scenario for a large Debye energy $E_D$ and for temperatures $E_F<T<E_D$, a mostly unexplored parameter regime also known as the semiquantum regime of transport~\cite{Spivak2006}.
Given some mild assumptions we indeed find an expression reminiscent of Eq.~(\ref{eq:planckformula}) in this regime. 
For this reason, we suggest to search for genuine Planckian transport at temperatures $E_F<T<E_D$ where the strongly interacting liquid itself is non-degenerate and the momentum is dissipated into a weakly coupled degenerate bath.
For comparison, at lower temperatures the degenerate electron liquid is found to not conform to the proposed form of Eq.~(\ref{eq:planckformula}).

It is important to note that a linear temperature dependence due to hydrodynamic effects has been put forward a long time ago for thermal transport in three dimensional neutral liquids~\cite{Andreev1978,Andreev1979}. However, for charge transport the relation between thermodynamic quantities and transport coefficients is qualitatively dissimilar due to screening~\cite{Andreev2011,Hartnoll2015}. 

The remainder of the paper is structured as follows. In Sec.~\ref{sec:II} we give a summarized account of the chain of arguments necessary to evaluate the hydrodynamic drag. We focus on two temperature ranges, respectively known as the semiclassical and the semiquantum regime of the electron liquid. The details of the calculation are documented in Sec.~\ref{sec:III}

\section{Summary of results} 
\label{sec:II}
Coulomb drag in a bilayer system is a well studied, sensitive probe for electronic correlations~\cite{Narozhny2016}.
The two layers contain electrons of effective mass $m_i$ and density $n_i$. For definiteness, the active layer obeys a generic quadratic dispersion with a Fermi energy $E_F$, Fermi velocity $v_F$; at high temperatures the average thermal velocity is $v_1(T)$. 
The Wigner crystal supports acoustic phonons with dispersion $\omega_q=\sqrt{v_s^2 q^2+\omega_0^2}$, where $v_s$ is the phonon velocity and $\omega_0\rightarrow 0$ the pinning potential, which we set to zero in the end. 
It was shown previously for such a setup that a small density of impurity sites yields a finite drag resistivity which is independent of disorder, up to logarithms~\cite{Braude2001}. 
It is therefore justified to assume $\tau_{dis}$ to be the largest timescale and still posit a steady state solution for the kinetic equation. 
The screening wavevectors are $\kappa_i=2\pi\nu_ie^2/\epsilon_i$,  where $\nu_i$ is the density of states at $T=0$ and $\epsilon_i$ is the dielectric constant in each layer $1$ and $2$. 

\subsection{Boltzmann-Langevin formalism}
Throughout, we set $\hbar=k_B=1$.
If the mean free path in the active layer is short compared to the layer separation ($v_1\tau_{eq}<d$), Coulomb drag is dominated by  collective (hydrodynamic) modes. 
It was recently argued that the drag resistivity is then given by an RPA-like expression~\cite{Chen2015,Kamenev1995,Flensberg1995}
\begin{align}
\frac{\rho_D}{\rho_Q}&=
\frac{\beta}{2^4 n_1n_2}
\int\dd \omega \int_0^{\infty} \frac{\dd q q^3 e^{-2q d}}
{\sinh^2 (\beta\omega/2)}
\notag\\&\quad\times
\frac{\Im \Pi_1^{-1}\Im \Pi_2^{-1}}
{\left|\frac{1-e^{-2qd}}{|V_0|^{-1}}-(\Pi_1^{-1}+\Pi_2^{-1})
+|V_0|^{-1}\Pi_1^{-1}\Pi_2^{-1}\right|^2},
\label{eq:drag}
\end{align}
where $\beta=T^{-1}$ is the inverse temperature, $\Pi_i$ is the charge susceptibility in each respective layer and $\rho_Q=e^2/2\pi$. The  bare Coulomb interaction is $V_0=-2\pi e^2/q\epsilon$.
While expression Eq.~\eqref{eq:drag} is perturbative in the interlayer Coulomb interaction,
the quasiparticle scattering rate which enters the kinetic equation can be large. 
Additionally, there is no restriction for the static screening length.
In the following we use Eq.~\eqref{eq:drag} to calculate the interlayer resistivity for intermediate temperatures. Compared to examples of hydrodynamic drag in symmetric setups~\cite{Apostolov2014,Patel2017}, the present system is in a regime where $\Im \Pi_1^{-1}\gg\Im \Pi_2^{-1}$, which to our knowledge has not been investigated previously.

The charge susceptibility relates fluctuations in the electric potential $\phi$ to charge fluctuations $\delta n=-e\Pi \phi $.
In the long wavelength limit the susceptibility in the active layer $1$ is uniquely determined by imposing charge, momentum and energy conservation. To this end, we split the nonequilibrium part  of the distribution function into two pieces, the quickly decaying fluctuations $\delta f$ not protected by any conservation laws and a slow, diffusive part $f_h$. The collision integral can likewise be decomposed. Firstly there are the frequent intralayer collisions ($\tau_{ee}^{-1}$), which respect charge, momentum and energy conservation but shuffle energy and momentum between single-particle excitations. Secondly, infrequent interlayer scattering ($\tau_{D}^{-1}$) move momentum and energy from one layer to the other. 
This scale separation $\tau_{ee}\ll \tau_D$ guarantees that $f_h$ decays much slower than $\delta f$ in layer $1$. For the same reason, the microscopic lifetime $\tau_{eq}=\tau_{ee}$ is known. In the relaxation time approximation, the linearized collision integral for the active layer thus becomes $I_1(f_{\bm{p}})\approx-\delta f/\tau_{ee}$.

\subsection{Coupling to electronic vibrations}
In the passive layer $2$ the momentum is stored in the form of electronic lattice vibrations. The momentum is scattered back into the first layer with the same rate $\tau_{D}^{-1}$. Additionally it relaxes due to extrinsic effects with rate $\tau_{dis}^{-1}$, which will be neglected when possible. 

Instead of using susceptibilities, the electronic response at intermediate temperatures has sometimes been discussed in terms of a semi-classical scattering cross-section~\cite{Friedrich2013}. 
However, even at the highest temperatures, far above the regime studied here, a two-dimensional electron gas does not approach a classical limit (cf. Sec.~\ref{app:classical}). For this reason, the hydrodynamic formulation with collective modes seems more appropriate.

The role of collective modes in the hydrodynamic regime has been thoroughly examined~\cite{Chen2015,Lucas2018a}, with the conclusion that a plasmon pole approximation can indeed capture the electronic response at fast frequencies. 
By enforcing conservation laws at short times, we find for temperatures $ E_F \beta\ll 1$
\begin{align}
\Pi_{1}&=\nu_1\left(\frac{m_1 \omega^2}{E_Fq^2}-\frac{1}{\alpha}\frac{2-3i\omega \tau_{ee}}{1-i\omega\tau_{ee}}\right)^{-1}.
\label{eq:pi1}
\end{align}
where $\alpha=E_F\beta$.
This susceptibility captures both the decrease of static screening with increasing temperatures and the enhanced damping rate of  collective charge fluctuations. At low temperatures ($\beta E_F\gg 1$), the intralayer susceptibility retains the same form, but with $\alpha=2$~\cite{Chen2015}.

The damping term which appears in Eq.~(\ref{eq:pi1}) is qualitatively different for $\omega \tau_{ee} \gtrless 1$. Depending on the dominant frequency cutoff for $\omega$ in the drag resistivity [Eq.~(\ref{eq:drag})], either case of this inequality may be realized.

The pinned electron lattice behaves like a clean and stiff dielectric with a long lifetime of the lattice excitations. We thus take for the charge susceptibility the standard form
\begin{align}
\Pi_2&=\nu_2v_s^2\left(\frac{(\omega+i\delta/2)^2-\omega_q^2}{q^2}\right)^{-1},
\end{align}
where $\delta=1/2\tau_D$. The coefficient is chosen such that in the absence of pinning, the static susceptibility matches the electron liquid, $\Re\Pi_2(\omega=\omega_0=0)=\nu_2$. More sophisticated response functions can easily be included, but they do not change the basic physics at play here.

\subsection{Semiquantum regime} 
\begin{figure}
\includegraphics[height=.49\columnwidth]{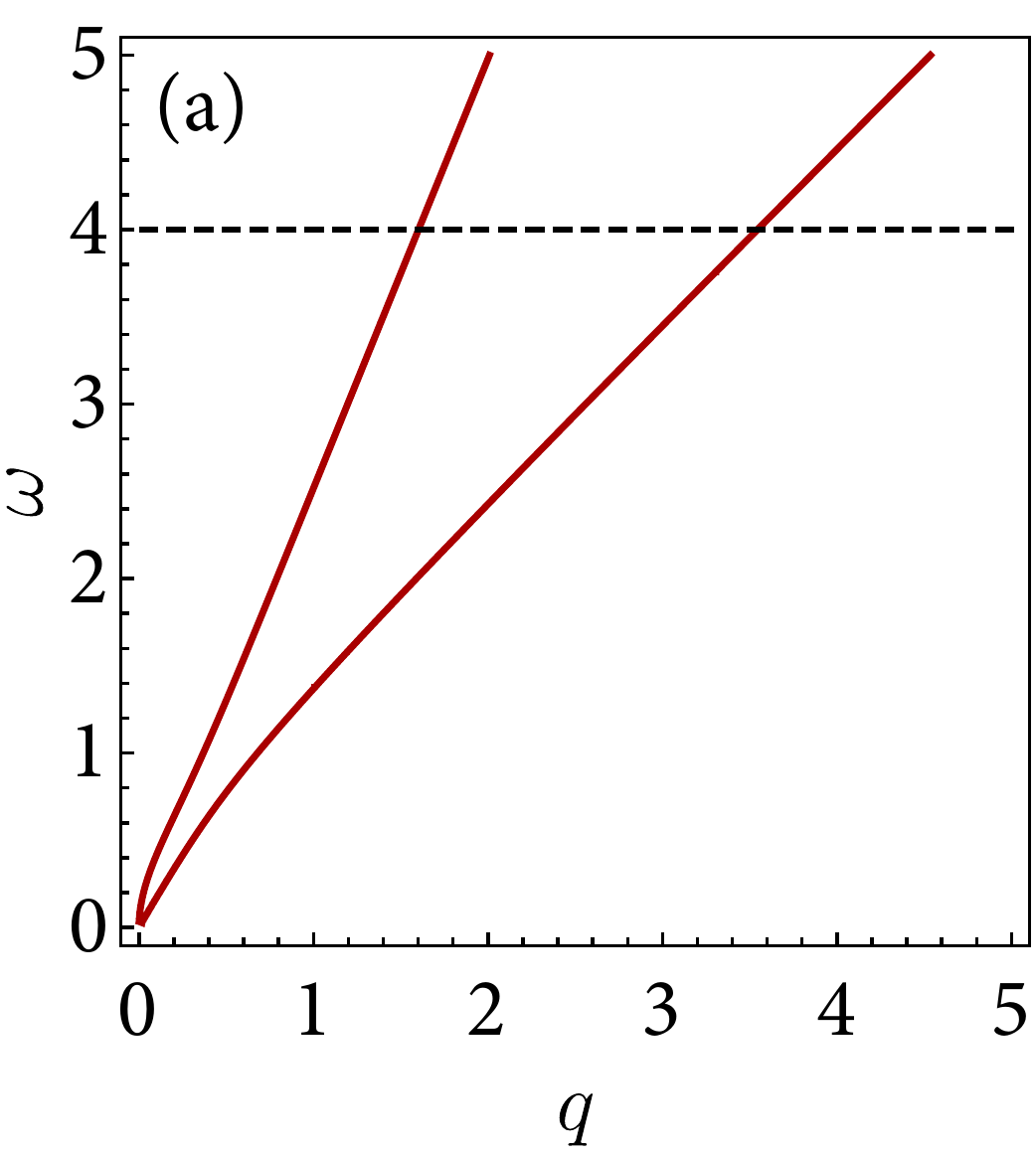}
\includegraphics[height=.49\columnwidth]{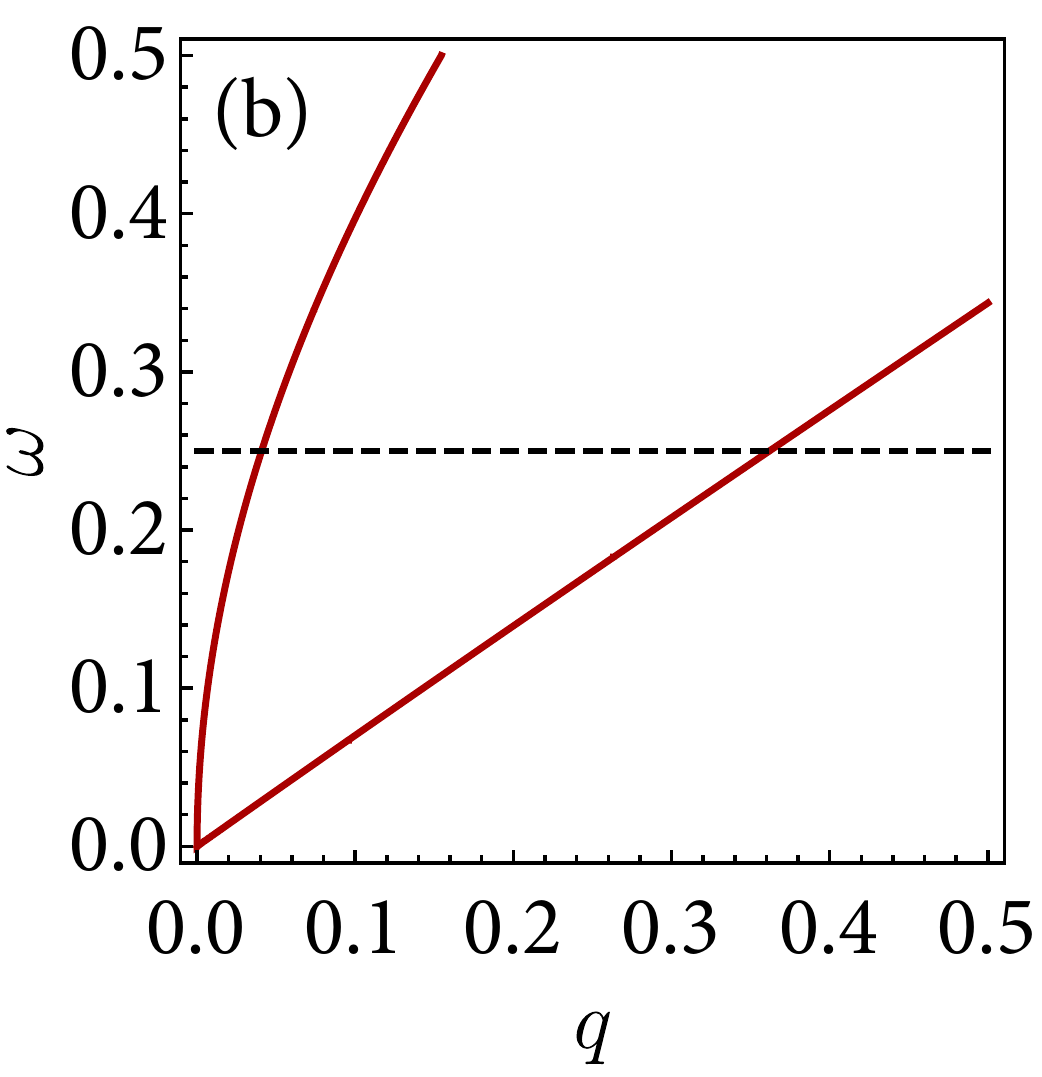}

\caption{Poles of the interlayer susceptibility in the $q$-$\omega$ plane. (a) If $T>E_F$, the dispersions of the pole lines are linear at the frequency cutoff. (b) For $T<E_F$, the dispersion is instead dominated by the small-$q$ part. The dashed black line is the frequency cutoff. Parameters are $\tau_{ee}=1$, $v_s=v_F=1$, $\kappa_1=\kappa_2=1$ and $\beta=0.25$ ($\beta=4$) for (a) and (b), respectively.}
\label{fig:poles}
\end{figure}

To extract the intrinsic hydrodynamic limit, we assume the interlayer separation to be larger than the mean free path $v_1\tau_{ee}$ in the active layer, but smaller than the coherence length $v_s\beta$ of the phonons in the passive layer. Importantly, there is always a range of distances where this is the case as long as $v_s>v_1$ because $\tau_{ee}$ is at least of size $\beta$.
This range of parameters can also be viewed as the expansion of the drag formula when screening is ineffective ($\kappa d\ll 1$), which is a result of the low electron density.
The location of the poles which appear as the zeros of the denominator in Eq.~(\ref{eq:drag}) depend on a few parameters, the quasiparticle velocity $v_1$, the phonon velocity $v_s$ and for high temperatures on $\beta$ (Fig.~\ref{fig:poles}). Importantly, if the phonon velocity is $v_s>v_1$, the poles are approximately located at $\omega=v_s q$ and $\omega=v_1 q$. 

Inserting the respective susceptibilities into Eq.~(\ref{eq:drag}) yields in the limit $\beta E_F \ll 1$
\begin{align}
\frac{\rho_D}{\rho_Q}&=
\frac{\pi}{72}
\frac{\kappa_1\kappa_2E_F}{n_1n_2v_s^2}\frac{1}{\tau_{ee}}
\log\left(1+\frac{9\tau_{ee}^2v_s^2}{4\lambda^2}\right).
\label{eq:dragEFbetasmall}
\end{align}
Here, $\lambda=\max(k_D^{-1},d,v_s\beta)$. For $E_F<T<E_D$, $\lambda=v_s\beta$ and the logarithm contains a large argument so that
\begin{align}
\frac{\rho_D}{\rho_Q}&=
\frac{\pi}{36}
\frac{\kappa_1\kappa_2E_F}{n_1n_2v_s^2}\frac{1}{\tau_{ee}}
\log\left(\frac{\tau_{ee}}{\beta}\right).
\label{eq:hydrodrag1}
\end{align}

We emphasize that the upper limiting temperature for this regime is not set by the Debye energy as in weakly interacting systems but by the condition $E_D\approx\tau_{ee}^{-1}$, which depends on the microscopic properties of both layers together. 

In a regime of Planckian intralayer scattering rates ($\tau_{ee}\sim \beta$), Eq.~(\ref{eq:dragEFbetasmall}) yields a resistivity linear in $T$, without logarithmic corrections. Taking the simplifying assumption that the electronic density is the same for both layers, which is possible by tuning the dielectric properties, the result closely resembles Eq.~(\ref{eq:planckformula}),
\begin{align}
\frac{\rho_D}{\rho_Q}&=
\mathcal{C}\frac{1}{\nu_1}
\frac{r_s^2}{v_s^2}\frac{1}{\tau_{ee}}.
\end{align}
The coefficient $\mathcal{C}$ is not universal, it takes a value $\mathcal{C}=\frac{\pi^2}{18}$ given identical dielectric constants in both layers. 
Remarkably, we recover the $T=0$ static charge susceptibility $\nu_1$ and a renormalized velocity $v_s/r_s$. 
Due to this effective diffusion velocity $v_s/r_s$, the drag resistivity is not  bounded for electron liquids with large $r_s$, but can increase with arbitrary slope. The latter is a result of the perturbative approach with regards to the interlayer interaction and breaks down once $v_s/r_s\sim v_1$.

\subsection{Semiclassical regime}
We now turn to the more familiar situation where $E_D<E_F$.
For a simple electron liquid, this condition is compatible with $v_s\ll v_F$. Further taking the screening wavevectors $\kappa_1$ and $\kappa_2$ to be smaller than the momentum cutoffs and the phonon scattering rate $\delta$ to be small compared to the frequency cutoff, the integration yields
\begin{align}
\frac{\rho_D}{\rho_Q}&=\frac{\pi}{16}
\frac{\sqrt{\kappa_1\kappa_2}v_s m_1}{n_1n_2\beta E_F}
\frac{\sqrt{\delta\tau_{ee}}\lambda^{-2}}{1+\sqrt{1+\tau_{ee}^2E_F/m_1\lambda^2}}
\label{eq:smallTresult}
\end{align}
with the same distance parameter $\lambda$ as before. 
We point out that the simplified result of Eq.~(\ref{eq:smallTresult}) used here for clarity contains all the essential ingredients which carry over to the more complicated case where the intricate parameter dependences are kept. 

Remarkably, the phonon scattering rate $\delta$ enters explicitly in Eq.~(\ref{eq:smallTresult}), a result of the smaller momentum space available at low temperatures. 
As discussed earlier, this phonon lifetime is bounded by the interlayer scattering time $\tau_D$, which itself enters in the definition of $\rho_D$. 
We therefore need to calculate this lifetime self-consistently from Eq.~(\ref{eq:smallTresult}). 
In the disorder dominated case, the momentum relaxation rate $\tau_D^{-1}$ would simply serve as the definition of the Drude resistivity $\rho_D$, making their relationship straightforward. 
While this is no longer necessarily true in the hydrodynamic limit, for the small driving current discussed here we can at least expect a laminar flow.
Using a suitably defined interlayer mass parameter $m_D$ and density $n_D$, which depend on the microscopic details of the system, the drag resistivity can then be written as $\rho_D=m_D/e^2 n_D \tau_D$. 
Given this approximation, the self-consistent solution is straightforward, with the result for $\tau_{ee}^2 E_F/m_1\ll \lambda^2$ being
\begin{align}
\frac{\rho_D}{\rho_Q}&= \frac{\pi^3}{2^8}
\frac{n_D\kappa_1\kappa_2m_1^2v_s^2 k_D^4}{n_1^2n_2^2m_DE_F^2}
\frac{\tau_{ee}}{\beta^2},
\label{eq:dragdegenerate}
\end{align}
In this result, the temperature only enters in the combination $\tau_{ee}/\beta^2$, 
making the hydrodynamic drag resistivity a sensitive probe for the intralayer electron-electron scattering rate. In particular, in a Fermi liquid where the quasiparticle lifetime follows $\tau_{ee}\sim \beta^2$ we obtain resistivity saturation for the drag. 
Importantly, the crossover to the highly correlated flow happens not precisely at temperature $E_D$ but at $\tau_{ee} v_F k_D\ll 1$, a combined quantity which contains properties of both drag layers. This threshold is identical with the Mott-Ioffe-Regel limit for the scattering time~\cite{Ioffe1960}.
Conversely, under the assumption that the electron liquid is constrained by the Planckian bound at these intermediate temperatures, one can  
insert a microscopic quasiparticle lifetime $\tau_{ee}\sim \beta$ into Eq.~(\ref{eq:dragdegenerate}), which would result in a linear-$T$ dependence of the drag resistivity. This is clearly reminiscient of the phenomenology of bad metals~\cite{Gunnarsson2003,Bruin2013}.
However, contrary to the semiquantum regime discussed before, the resistivity is no longer proportional to the microscopic scattering rate as proposed in Eq.~(\ref{eq:planckformula}), 
but instead proportional to its inverse, the scattering time. 
We conclude therefore that the momentum diffusion constant entering the drag resistivity cannot be associated with a form $D=v^2\tau_{eq}$ such that the right hand side has an obvious microscopic interpretation.

Since $\rho_D\sim \tau_{ee}$, increasing the electron-electron interaction will reduce the drag resistivity, 
which demonstrates how strong but momentum-conserving interactions can result in a scale separation of energy and momentum relaxation rates without invoking umklapp scattering. 
In this scenario, the resistivity will contain an unusually small $T^2$-dependence compared to the leading coefficient of the thermal resistivity originating from the much faster quasiparticle relaxation rate. 
This entails a reduced Lorenz number and a deviation from the Wiedemann-Franz law~\cite{Mahajan2013a,Jaoui2018}. 

We conclude that hydrodynamic drag in the semiclassical regime closely resembles the phenomenology which is expected for charge transport in strongly correlated electron systems, but seemingly with the 'wrong' dependence on the microscopic timescale. 

\section{Electron susceptibility}
\label{sec:III}
Hydrodynamic drag is by definition dominated by the collective modes of the Coulomb liquid and the Wigner crystal in layer $1$ and $2$, respectively. 
In the following we explain the ingredients needed for a calculation of intralayer and interlayer susceptibilities in this regime using the Boltzmann-Langevin approach.

As mentioned previously, we concentrate on a simple quadratic dispersion amenable to a controlled calculation. However, the same results are expected to hold in the general case, except for numerical coefficients. This is related to the very large temperatures $T\sim E_F$ that we assume for the most part of the calculation, which will average out fine details of the dispersion.
On the flip side, at temperatures $T\sim E_F$, in the distribution function the temperature dependence of the chemical potential has to be taken into account. For a dispersion $\epsilon(\bm{k})=k^2/2m-\mu=k^2/2m-\mu(\beta)$ and fixed electron density $n=k_F^2/2\pi$, the chemical potential $\mu$ is determined by the condition
\begin{align}
n&=\frac{2}{2\pi}\int_0^\infty\frac{k\,\dd k}{1+e^{\beta (k^2/2m-\mu)}}\\
&=\frac{\log(1+e^{\beta \mu})}{\pi\beta/m}\\
e^{\beta \mu}&=e^{\beta E_F}-1,
\end{align}
where $E_F=\pi n/m$ and the factor $2$ accounts for both spin species. 

\subsection{Intralayer susceptibility}
The calculation of the susceptibility follows along the lines of~\cite{Chen2015}. The kinetic equation is given by
\begin{align}
&(i\bm{q}\cdot\bm{v}-i\omega)\Delta f(\bm{q},\bm{p},\omega)
+i e \phi(\bm{q},\omega) \bm{q}\cdot\bm{v}\partial_{\epsilon} f(\bm{q},\bm{p},\omega)
\notag\\
&\quad=-\frac{\Delta f(\bm{q},\bm{p},\omega)-f_h(\bm{q},\bm{p},\omega)}{\tau_{ee}}.
\end{align}
Here, $f(\bm{q},\bm{p},\omega)$ is the distribution function in Fourier representation and $\phi(\bm{q},\omega)$ is the electrical potential.
For sake of brevity, let $\tau_{ee}=\tau$ and $f(\bm{q},\bm{p},\omega)=f$. As explained previously, the complete non-equilibrium part of the distribution function is $\Delta f$, the long-lived modes are $f_h$ and the fast decaying part is simply the difference of the two.
We extract the hydrodynamic part of the susceptibility at $ E_F\beta \ll 1$ by imposing  particle, momentum and energy conservation for the collision integral at short times,
\begin{align}
\int \frac{\dd ^2 p }{(2\pi)^2} (\Delta f-f_h)=0
\label{eq:cons1}\\
\int \frac{\dd ^2 p }{(2\pi)^2} \bm{p}(\Delta f-f_h)=0
\label{eq:cons2}\\
\int \frac{\dd ^2 p }{(2\pi)^2} (\epsilon-\mu)(\Delta f-f_h)=0.
\label{eq:cons3}
\end{align}
The long lived non-equilibrium part of the distribution function $f_h$ correspondingly carries three modes,
\begin{align}
f_h&=-\partial_{\epsilon} f
\left(\rho^{-1}\delta n+m \bm{u}\cdot\bm{v}
+\beta \delta T(\epsilon-\mu)\right),
\end{align}
where $\delta n$ are charge fluctuations, $\bm{u}$ the flow velocity and $\delta T$ temperature fluctuations.
The non-equilibrium distribution function $\Delta f$ is determined from the kinetic equation
\begin{align}
\Delta f&=
\frac{\tau^{-1}f_h-i e \phi \bm{q}\cdot\bm{v} 
\partial_{\epsilon} f}
{i\bm{q}\cdot\bm{v}-i\omega+\tau^{-1}}.
\end{align}
Inserting these definitions into Eq.~(\ref{eq:cons1}-\ref{eq:cons3}) the resulting system of equations is
\begin{align}
I_0\delta n + \beta E_1 \rho\delta T
&= A_0 \rho e\phi +\delta n\Delta_0
+\beta\Delta_1\rho\delta T
\notag\\&\quad
-i\frac{A_0}{q^2\tau}\rho m \bm{u}\bm{q}\\
n\bm{u}\bm{q}&=i\frac{1-i\omega\tau}{\tau}A_0\rho e \phi 
-i\frac{A_0}{\tau}\delta n
-i\frac{\beta}{\tau}A_1\rho\delta T
\notag\\&\quad
+\frac{1-i\omega\tau}{q^2\tau^2}A_0\rho m\bm{u}\bm{q}
\\
E_1\delta n +
\beta E_2\rho \delta T&=
A_1\rho e\phi
+\Delta_1\delta n
+\beta\Delta_2\rho\delta T
\notag\\&\quad 
-i\frac{A_1}{q^2\tau}\rho m \bm{u}\bm{q}
\end{align}
Here, we introduced the following shorthands
\begin{align}
\Delta_0&=
\int_0^{\infty}\dd \epsilon
(-\partial_{\epsilon}f)  D\\
\Delta_1&=
\int_0^{\infty}\dd \epsilon
(-\partial_{\epsilon}f) (\epsilon-\mu)D\\
\Delta_2&=
\int_0^{\infty}\dd \epsilon
(-\partial_{\epsilon}f) (\epsilon-\mu)^2 D
\shortintertext{where}
D^{-1}&=\sqrt{(1-i\omega\tau)^2+q^2\tau^2\frac{2\epsilon}{m}}
\end{align}
For a quadratic dispersion the energy integrals can be performed explicitly,
\begin{align}
I_n&=-\beta^{-n}Li_n(1-e^{\beta E_F})\\
A_0&=I_0-(1-i\omega\tau)\Delta_0\\
A_1&=I_1-I_0\mu-(1-i\omega\tau)\Delta_1\\
E_1&=I_1-I_0\mu\\
E_2&=I_2-2\mu I_1+I_0\mu^2
\end{align}
We expand $\Delta_i$ up to $\mathcal{O}(q^6)$ and $I_i$ to order $\mathcal{O}(\beta)$, which corresponds to a number of $i+1$ terms in $I_i$.
Finally, the susceptibility is given by $\delta n=-\Pi e \phi $, where $\delta n$ fulfills the continuity equation $\delta n +\mathbf{q}\mathbf{u}=0$. In contrast to low temperatures, $\nu \delta\mu\neq \delta n $. 
This yields Eq.~(\ref{eq:pi1}),
\begin{align}
\Pi_{1}&=\nu_1\left(\frac{m_1 \omega^2}{E_Fq^2}-\frac{1}{\alpha}\frac{2-3i\omega \tau_{ee}}{1-i\omega\tau_{ee}}\right)^{-1}.
\end{align}

\subsection{Interlayer susceptibility}
The Coulomb drag is to a good approximation given by the perturbative result for the interlayer susceptibility, Eq.~\eqref{eq:drag}.
To resolve the desired slow end of the excitation spectrum, i.e. the hydrodynamic couling, it is appropriate to expand in $\beta\omega\ll1$. We then make use of the Thomas-Fermi screening wavevectors $\kappa_i=2\pi\nu_ie^2/\epsilon_i$ to rewrite  the drag formula of Eq.~(\ref{eq:drag}) using dimensionless susceptibilities $\bar\Pi_i=\Pi_i/\nu_i$
\begin{widetext}
\begin{align}
\frac{\rho_D}{\rho_Q}&=
\frac{\kappa_1\kappa_2}{4 n_1n_2\beta}
\int\dd \omega \int_0^{\infty} \frac{\dd q q^3}
{\omega^2}
%\notag\\&\quad\times
\frac{\Im \bar\Pi_1^{-1}\Im \bar\Pi_2^{-1}}
{\left|\sinh(qd)\tfrac{\kappa_1\kappa_2}{q}-e^{q d}(\kappa_2\bar\Pi_1^{-1}+\kappa_1\bar\Pi_2^{-1})
+e^{q d}q\bar\Pi_1^{-1}\bar\Pi_2^{-1}\right|^2}.
\label{eq:dragspecific}
\end{align}
\end{widetext}
In the denominator, the first term is subleading for the weak screening limit $\kappa d\to 0$ and can be dropped. It is then a the cutoff scale which determines which of the remaining terms become dominant. Crucially, it can happen that the second term, while formally subleading, acts as a regulator for the phase space integral. In this case, in the clean limit the interlayer susceptibility can acquire a resonance mismatch where the  numerator is zero at the zeros of the denominator.
For example, in the limit that $\Im \Pi_2^{-1}/\Im \Pi_1^{-1}\rightarrow 0$, the drag will vanish unless the denominator factorizes and provides a Lorentz peak that renders the integral finite. And indeed, in the semiquantum case the denominator factorizes, but for a degenerate electron liquid it does not. Therefore, the latter case requires a self-consistent evaluation of the drag formula.

\subsection{Semiquantum Coulomb liquid}
We evaluate Eq.~\eqref{eq:dragspecific} for temperatures $ E_F < T < E_D$.
To repeat, the dimensionless susceptibilities are 
\begin{align}
\bar\Pi_{1}^{-1}&=\frac{m_1 \omega^2}{E_Fq^2}-\frac{1}{E_F\beta}\frac{2-3i\omega \tau}{1-i\omega\tau}\\
\bar\Pi_2^{-1}&=\frac{\omega^2-\omega_q^2}{v_s^2q^2}
+\frac{i\omega\delta}{v_s^2q^2}.
\end{align}
Since $E_F\beta\ll 1$, this simplifies to
\begin{align}
\frac{\rho_D}{\rho_Q}&=
\frac{\kappa_1\kappa_2}{4 n_1n_2\beta}
\int\dd \omega \int_0^{\infty} 
\frac{q\dd q }{E_F\beta v_s^2}
\frac{\tau\delta}{1+\omega^2\tau^2}
\notag\\&\quad
\left|\frac{1}{E_F\beta}
\frac{2-3i\omega \tau}{1-i\omega\tau}\right|^{-2}
\left|-\kappa_2+q\left(\frac{\omega^2-\omega_q^2}{v_s^2q^2}
+\frac{i\omega\delta}{v_s^2q^2}\right)\right|^{-2}
\end{align}
For small $\delta$ the frequency integral is dominated by the Lorentz peak in the last term, which yields
\begin{align}
&=\frac{\pi}{4}\frac{\kappa_1\kappa_2 E_F}{n_1n_2v_s^2}
\int_0^{\infty} \dd q
\frac{\tau}{(1+\omega_q^2\tau^2)}
\notag\\&\quad
\frac{q \tau}{4+9\tau^2(\kappa_2+q)^2\omega_q^2/q^2}
\frac{\omega_q^2}{(\kappa_2+q)^2}.
\end{align}
We can safely neglect the screening wavevector in the denominators, because the integral is dominated by the behavior at large momenta.
The $q$ integral is cut once $\omega_q$ reaches the frequency cutoff, at which point the integration no longer reaches the location of the poles. Introducing the short distance cutoff scale $\lambda$, we arrive at Eq.~(\ref{eq:dragEFbetasmall}).
%\begin{align}
%\frac{\rho_D}{\rho_Q}&=
%\frac{\pi}{72}\frac{\kappa_1\kappa_2E_F}{n_1n_2v_s^2}
%\frac{1}{\tau_{ee}}
%\log\left(1+\frac{9\tau_{ee}^2v_s^2}{4\lambda^2}\right).
%\end{align}

\subsection{Degenerate Coulomb liquid}
For the degenerate Coulomb liquid and intermediate but not too small temperatures $E_D < T < E_F$, the susceptibility was calculated by \citeauthor{Chen2015}~\cite{Chen2015}
\begin{align}
\bar\Pi_{1}^{-1}&=
\frac{m \omega^2}{E_Fq^2}
-\frac{2-3i\omega \tau}{2(1-i\omega\tau)}
\notag\\&\approx
\frac{m \omega^2}{E_Fq^2}
-1+\frac{i\omega \tau}{2(1+\omega^2\tau^2)}
\end{align}
The denominator in Eq.~(\ref{eq:dragspecific}) can be written more explicitly with real and imaginary parts of the inverse susceptbilities $\bar\Pi_n^{-1}=R_n+i I_n$ as
\begin{align}
&(\kappa_2 R_1 + \kappa_1 R2 - q R_1 R_2)^2
+ I_1^2 (\kappa_2 - q R_2)^2 
+ 2\kappa_1 \kappa_2 I_1 I_2
\notag\\&\quad
+ I_2^2(\kappa_1 - q R_1)^2 
+q^2 I_1^2 I_2^2.
\label{eq:denoms}
\end{align}
Since the dissipative processes in the second layer are slow it suffices to keep the linear term in $I_2$ in Eq.~(\ref{eq:denoms}). Since it only has minor impact numerically, we also leave out higher powers in $I_1$, to obtain for the denominator (cf. Sec.~\ref{app:B})
\begin{align}
\frac{m^2}{q^2\omega_q^4E_F^2}
(\omega^2-\omega_+^2)^2(\omega^2-\omega_-^2)^2
+\frac{\kappa_1\kappa_2 \tau}{\omega_q^2(1+\omega^2\tau^2)}\omega^2\delta
\end{align}
with the roots given by
\begin{align}
2q\omega_{\pm}^2&=
q^2(q+\kappa_1)\tfrac{E_F}{m}+\omega_q^2(q+\kappa_2)
\mp\biggl[(q^2(q + \kappa_1)\tfrac{E_F}{m} 
\notag\\&\quad
+\omega_q^2 (q +\kappa_2))^2-4q^3\omega_q^2(q +\kappa_1+\kappa_2)\tfrac{E_F}{m}\biggr]^{1/2}.
\end{align}
We use the approximation that for small $\delta$
\begin{align}
&\int\dd\omega \frac{f(\omega)\delta}{(\omega^2-\omega_+^2)^2(\omega^2-\omega_-^2)^2+A(\omega)^2\omega^2\delta}
\notag\\
&\approx\frac{\pi}{2}\frac{\sqrt{\delta}}
{|\omega_+^2-\omega_-^2|}\left(\frac{f(\omega_+)}{A(\omega_+)\omega_+^2}+
\frac{f(\omega_-)}{A(\omega_-)\omega_-^2}\right),
\end{align}
where the smooth $\omega$ dependences are kept only through the replacement $\omega\rightarrow\omega_{\pm}$. The drag is then
\begin{align}
\frac{\rho_D}{\rho_Q}&=
\frac{\pi}{16}
\frac{\sqrt{\kappa_1\kappa_2}E_F}{n_1n_2\beta m_1}
\int_0^{\infty} \frac{\dd q q^4 \omega_q\sqrt{\tau \delta}}
{|\omega_+^2-\omega_-^2|}
\notag\\&\quad\times
\left(\frac{1}{\sqrt{1+\omega_+^2\tau^2}\omega_+^2}+
\frac{1}{\sqrt{1+\omega_-^2\tau^2}\omega_-^2}\right).
\end{align}
This expression does not contain further easy simplifications unless the parameter regime is severely restricted. Anticipating that the important contributions come from the cutoff region, one can set $\kappa_1=\kappa_2=0$ in the denominators. In line with the desired temperature range $E_D<T<E_F$, we can further assume that $v_s^2\ll E_F/m_1$. This only leaves
\begin{align}
\frac{\rho_D}{\rho_Q}&=
\frac{\pi}{16}
\frac{\sqrt{\kappa_1\kappa_2 }m_1}{n_1n_2\beta E_F}
\int_0^{\infty} \dd q\frac{ \sqrt{\tau \delta}v_s q}
{\sqrt{1+q^2\tau^2 E_F/m}},
\end{align}
which immediately yields Eq.~(\ref{eq:smallTresult}) when using the large momentum cutoff $q\sim 1/\lambda$. We note that the more complete expression Eq.~(\ref{eq:dragspecific}) leads to the same self-consistent solution, only with more involved coefficients.

\subsection{Connection to the ohmic limit}
The previously presented regimes have not been discussed before in the literature for a bilayer system composed of an electron liquid and a Wigner crystal. However, some results exist for the low temperature limit~\cite{Braude2001}. We now show that our results connect to these earlier findings.
For the lowest temperatures, the inverse dimensionless susceptibility of the electron liquid becomes~\cite{Apostolov2014}
\begin{align}
\bar\Pi_{1}^{-1}&=\left(1-\frac{\omega}{\sqrt{\omega^2-v^2q^2}}\right)^{-1}
\end{align}
To make the scaling easier we only keep the cutoff in $\omega$ and refrain from integrating the $\sinh$ in the drag formula explicitly. This approximation only affects the coefficients of the result, not the parametric dependencies.
For the same reason, we set $\kappa_1=\kappa_2=\kappa$.
As before, the drag integral has several cutoffs, but in any case it is true that $|\omega|< 1/\beta$ and $q<1/d$. The crossover between various intermediate regimes is governed by other cutoffs if they are stricter than these global ones. We note that at lowest $T$, the Debye frequency is not important. The three terms in the denominator of Eq.~\eqref{eq:dragspecific} have sizes $d\kappa^2$, $\kappa$ and $q$. The first term is thus dominant if $\kappa d>\min(1,q/\kappa)=1$. This is the far limit, drag is hydrodynamic and proportional to the phonon scattering rate $\delta$. 
\begin{align}
\frac{\rho_D}{\rho_Q}&\sim \frac{\delta}{\beta d^2}\min(1/d,1/v_s\beta)
\end{align}
The layers have little influence on each other and drag vanishes in the clean limit. On the other hand, if $\kappa d<1$, one can neglect the static term in the denominator,
\begin{align}
\frac{\rho_D}{\rho_Q}&=
\frac{\kappa_1\kappa_2}{4 n_1n_2\beta}
\int\dd \omega \int_0^{\infty} \frac{\dd q q^3}
{\omega^2}
\notag\\&\quad\times
\frac{\Im \bar\Pi_1^{-1}\Im \bar\Pi_2^{-1}}
{\left|-\kappa(\bar\Pi_1^{-1}+\bar\Pi_2^{-1})
+q\bar\Pi_1^{-1}\bar\Pi_2^{-1}\right|^2}.
\end{align}
This limit is the natural limit at low density, where the screening length diverges. Let us concentrate on the instances where $\delta$ is eliminated from the result. Roughly speaking, this happens either if $|\bar\Pi_1^{-1}| \ll |\bar\Pi_2^{-1}|$, or if $\kappa\ll q$. In the first case
\begin{align}
\frac{\rho_D}{\rho_Q}&=
\frac{\kappa_1\kappa_2}{4 n_1n_2\beta}
\int\dd \omega \int_0^{\infty} \frac{\dd q q^3}
{\omega^2}
\frac{\Im \bar\Pi_1^{-1}\Im \bar\Pi_2^{-1}}
{|\bar\Pi_2|^{-2}\left|
q\bar\Pi_1^{-1}-\kappa\right|^2}.
\end{align}
For $\kappa\gg q$, this yields the result reported before
\begin{align}
\frac{\rho_D}{\rho_Q}&\sim \frac{1}{\beta}\min(1/d,1/v_s\beta)^3.
\end{align}
which requires a Wigner crystal with a small susceptibility (e.g. due to being pinned). In particular, at the lowest temperatures the T-dependence of the drag resistivity is $\sim T^4$. Note that in this case no dependence on the quasiparticle lifetime enters the drag formula~\cite{Braude2001}. 
If however $\kappa \ll q$ the result
becomes
\begin{align}
\frac{\rho_D}{\rho_Q}&=
\frac{\kappa_1\kappa_2}{4 n_1n_2\beta}
\int\dd \omega \int_0^{\infty} \frac{\dd q q^3}
{\omega^2}
\frac{\Im \bar\Pi_1^{-1}\Im \bar\Pi_2^{-1}}
{\left|q\bar\Pi_1^{-1}\bar\Pi_2^{-1}\right|^2},
\end{align}
which integrates to
\begin{align}
\frac{\rho_D}{\rho_Q}&\sim \frac{1}{\beta}\min(1/d,1/v_s\beta)
\end{align}
This is the usual momentum loss due to soft phonons, which is of order $T^2$ at the lowest temperatures, and crosses over to $T^1$ once $d/v_s>\beta$. We reiterate that the latter linear-T dependence, unlike the one discussed for the strongly coupled temperature regimes, is independent of the quasiparticle lifetime.

\subsection{Scope of applicability}
Finally, we discuss some limitations of the hydrodynamic regime presented above.
By construction, the fluctuation induced momentum exchange between both layers does not induce a finite macroscopic velocity of the phonons in the passive layer. 
Not covered hereby is the extreme case where the second layer stores momentum essentially indefinitely, such a breakdown of the steady state occurs if the disorder mean free path exceeds the sample size. 

In the opposite case where the shortest timescale is due to disorder instead of interactions the collision integral relaxes all  fluctuations with the rate of disorder scattering. 
This results in the more familiar ballistic/ohmic behavior which we already explained. 

\section{Conclusions} 
We have suggested a Coulomb drag experiment using a Wigner crystal to sensitively diagnose the electron-electron scattering rate for strongly interacting, clean Coulomb liquids. 
For temperatures $E_D<T<E_F$, we predict resistivity saturation given conventional electron-electron dominated quasiparticle lifetimes and a linear-T resistivity proportional to the quasiparticle lifetime for the proposed Planckian regime of strong interactions.
In the semiquantum case with $E_F<T<E_D$, interlayer drag is instead proportional to the quasiparticle scattering rate, with an effective momentum diffusion constant which closely resembles the form obtained from holographic methods. 
In either situation, the momentum conserving electron-electron interaction becomes the relevant time scale for charge transport, but in a distinctly different fashion. 

The mechanism put forth here is not restricted to coupling to a Wigner crystal, the two-fluid construction with a definitive hierarchy of lifetimes might also play a role in a number of low density materials with small Fermi pockets, where the phonons can provide a highly coherent yet almost momentum conserving bath. A timely example are anisotropic Weyl semimetals, if their chemical potential is not too close to the critical point. 
For example, the charge transport in $\mathrm{WP}_2$ was measured to obey a Planckian scaling for an intermediate temperature regime~\cite{Gooth2017}, however we caution that it is presently unclear whether the relevant electron-electron processes are mostly two-dimensional.

The Coulomb drag geometry could also be useful to map out the temperature dependence of the quasiparticle lifetime over a much wider temperature regime, which was suggested to be non-monotonic~\cite{Bard2018}.

\begin{acknowledgments}
A. Stern was very helpful in the initial stages of the project. We also thank A. Levchenko, O. Golan, F. von Oppen, P. Ostrovsky and Y. Werman for helpful discussions. T. H. is supported by the Minerva Foundation.
\end{acknowledgments}

\appendix

\setcounter{equation}{0}
\renewcommand{\theequation}{A.\arabic{equation}}

\section{Comment on the semiclassical limit in 2D}
\label{app:classical}
We briefly comment on the absence of a proper classical limit for the scattering cross section of a 2DEG. This reinforces the proposal of the main text that it is worthwhile to search for the Planckian scattering regime at temperatures above the Fermi energy. To this end, we are interested in the two-dimensional motion of an electron with velocity $v$ and energy $E=mv^2/2$ subject to a Coulomb potential $V(r)=\frac{C}{r}$.
Upon elastic scattering from a static target the classical scattering angle $\theta$ obeys
\begin{align}
\tan \frac{\theta}{2}&=\frac{b_0}{b},
\label{deftheta}
\end{align}
where $b$ is the impact parameter and $b_0=C/2E$ is the length scale associated with a large scattering angle. 
The two-dimensional (differential) cross section is defined as
\begin{align}
\frac{\dd \sigma}{\dd \Omega}
&=\left|\frac{\dd b(\theta)}{\dd \theta}\right|.
\end{align}
This leads to a  cross section of~\cite{Barton1983}
\begin{align}
\frac{\dd \sigma_{2d,c}}{\dd \Omega}&=\frac{b_0}{2\sin^2\frac{\theta}{2}}.
\end{align}
For a density of scatterers $n$, a rough estimate for the scattering rate is~\cite%[p. 167]
{Goldston1995}
\begin{align}
\frac{1}{\tau_{2d}}&=\frac{n v}{2}\int_0^\infty \sin^2\theta(b)\dd b\displaybreak\\
&=nvb_0\int_0^\infty \frac{2x^2}{(1+x^2)^2}\dd x
=\frac{\pi}{2} n v b_0.
\label{coul2d}
\end{align}
This relation follows from the assumption of elastic scattering. The initial velocity $v$ is parallel to the coordinate axis and $v_\perp=0$. Therefore for a single particle trajectory it holds
\begin{align}
(\Delta v_\perp)^2&=v^2\sin^2\theta.
\end{align}
Integration over all incoming particles leads to
\begin{align}
\frac{\dd\langle(\Delta v_\perp)^2\rangle}{\dd t}
&=n v \int (\Delta v_\perp)^2 \dd b.
\end{align}
For elastic scattering energy conservation implies
\begin{align}
(v+\Delta v_{||})^2+(\Delta v_\perp)^2&=
v_{||}^2\\
2v(\Delta v_{||})+(\Delta v_\perp)^2+(\Delta v_{||})^2&=0,
\end{align}
therefore $v(\Delta v_{||})\sim (\Delta v_\perp)^2$ and to leading order
\begin{align}
-\frac{1}{v}
\frac{\dd\langle\Delta v_{||}\rangle}{\dd t}
&=
\frac{1}{2v^2}
\frac{\dd\langle(\Delta v_\perp)^2\rangle}{\dd t}
=\frac{1}{\tau_{2d}}.
\end{align}
By definition, for a classical calculation the length scale $b_0$ cannot become smaller than the de Broglie wavelength $\lambda_B=\hbar/mv$. This limitation acts as a high energy cutoff, which is not logarithmically weak, as it is the case for 3D, but multiplicative.
It is important to keep in mind that additionally, in 2d there is a second and unrelated effect besides this IR cutoff: The solution of Schr\"odinger's equation for Coulomb scattering changes the scattering cross section itself, yielding~\cite{Barton1983}
\begin{align}
\frac{\dd\sigma_{2d,q}}{\dd \Omega}&=\frac{\dd \sigma_{2d,c}}{\dd \Omega}\tanh\frac{\pi C}{\hbar v}.
\end{align}
With increasing temperature and thus increasing velocity $v$ these reentrant quantum effects $b_0\gg\lambda_B$ (i.~e. $C\gg \hbar v$) invalidate the classical limit.

To reiterate, we do not enter this regime in the main text by restricting the entire discussion to a strongly correlated system with a small quasiparticle lifetime, which is quintessentially different from the high-temperature limit. In the this case, a new type of hydrodynamic regime can be reached where momentum dissipation is governed by the timescale of microscopic, momentum conserving processes.

\section{Resonance mismatch}
\label{app:B}
Here, we give some further insights into the peculiar behavior of the interlayer susceptibility in the semiclassical regime, $E_d<T<E_F$.
When the poles in the interlayer susceptibility are off resonance, two scenarios can occur. To see this, we compare terms of size $\delta$ and $\delta^2$ in the denominator. At low temperatures, the linear term is more important than the quadratic one if
\begin{align}
\kappa^2 I_1 I_2&> I_2^2(\kappa-q R_1)^2+q^2 I_1^2I_2^2\\
\delta&> \frac{\kappa^2 I_1 v_s^2q^2}{\omega((\kappa-q R_1)^2+q^2I_2^2)}\\
\delta&>\frac{q}{\omega} 
\end{align}
This condition is violated in many places in the $q-\omega$ plane, even though it marginally holds at the poles of the phonon propagator. What this means is that terms of order $\delta$ to leading order are not entering in the calculation. Then, the $\delta^2$ dominates the pole placement and a Lorentz form emerges.

The low temperature approximation is inapplicable if the electron lifetime ceases to be much larger than the temperature, $\tau\omega\sim 1$, i.e. once we reach the strongly coupled regime of the electron liquid. In this case, the susceptibility is approximately
\begin{align}
\bar\Pi_{1}^{-1}&=
\frac{m \omega^2}{E_Fq^2}
-1+\frac{i\omega \tau}{2(1+\omega^2\tau^2)}
\end{align}
We once again compare terms of size $\delta$ and $\delta^2$ in the denominator:
\begin{align}
\kappa^2 I_1 I_2&> I_2(\kappa-q R_1)^2+q^2 I_1^2I_2^2\\
\delta&> \frac{\kappa^2 I_1 v_s^2q^2}{\omega((\kappa-q R_1)^2+q^2I_2^2)}\\
\delta&>\frac{1}{1+\omega^2\tau^2} 
\end{align}
This condition will become true around the upper cutoff of the integration. This means that a calculation using only the $\delta^2$ term receives an additional cutoff which reintroduces the linear-$\delta$ dependence in the calculation. Equivalently, one can say that an expansion of the drag in $\kappa$ results in a divergent integrand which needs a regulator. At any rate, the denominator is intrinsically detuned from a Lorentz form with respect to $\delta$. This implies immediately that the drag response vanishes in the limit $\delta\to 0$.

From yet another angle, one can say that at low temperatures, the imaginary part of the susceptibility is just the Landau damping term, which is of order 1, while at higher temperatures it is $\omega\tau/(1+\omega^2\tau^2)$ which peaks at $\sqrt{\tau}$, making it unrestricted. This enhances the mixed term and favors it over the higher order terms. The liquid becomes highly damped and can no longer  adequately respond to the resonance. Since both fluids are desynchronized, the phonons are then getting scattered as well, which makes $\delta$ an instrinsic quantity and introduces a feedback between the layers. For this reason, it is most approriate to keep the term linear in $\delta$ and seek a self-consistent evaluation of the integral.


\begin{thebibliography}{38}%
\makeatletter
\providecommand \@ifxundefined [1]{%
 \@ifx{#1\undefined}
}%
\providecommand \@ifnum [1]{%
 \ifnum #1\expandafter \@firstoftwo
 \else \expandafter \@secondoftwo
 \fi
}%
\providecommand \@ifx [1]{%
 \ifx #1\expandafter \@firstoftwo
 \else \expandafter \@secondoftwo
 \fi
}%
\providecommand \natexlab [1]{#1}%
\providecommand \enquote  [1]{``#1''}%
\providecommand \bibnamefont  [1]{#1}%
\providecommand \bibfnamefont [1]{#1}%
\providecommand \citenamefont [1]{#1}%
\providecommand \href@noop [0]{\@secondoftwo}%
\providecommand \href [0]{\begingroup \@sanitize@url \@href}%
\providecommand \@href[1]{\@@startlink{#1}\@@href}%
\providecommand \@@href[1]{\endgroup#1\@@endlink}%
\providecommand \@sanitize@url [0]{\catcode `\\12\catcode `\$12\catcode
  `\&12\catcode `\#12\catcode `\^12\catcode `\_12\catcode `\%12\relax}%
\providecommand \@@startlink[1]{}%
\providecommand \@@endlink[0]{}%
\providecommand \url  [0]{\begingroup\@sanitize@url \@url }%
\providecommand \@url [1]{\endgroup\@href {#1}{\urlprefix }}%
\providecommand \urlprefix  [0]{URL }%
\providecommand \Eprint [0]{\href }%
\providecommand \doibase [0]{http://dx.doi.org/}%
\providecommand \selectlanguage [0]{\@gobble}%
\providecommand \bibinfo  [0]{\@secondoftwo}%
\providecommand \bibfield  [0]{\@secondoftwo}%
\providecommand \translation [1]{[#1]}%
\providecommand \BibitemOpen [0]{}%
\providecommand \bibitemStop [0]{}%
\providecommand \bibitemNoStop [0]{.\EOS\space}%
\providecommand \EOS [0]{\spacefactor3000\relax}%
\providecommand \BibitemShut  [1]{\csname bibitem#1\endcsname}%
\let\auto@bib@innerbib\@empty
%</preamble>
\bibitem [{\citenamefont {{Son}}\ and\ \citenamefont
  {{Starinets}}(2007)}]{Son2007}%
  \BibitemOpen
  \bibfield  {author} {\bibinfo {author} {\bibfnamefont {D.~T.}\ \bibnamefont
  {{Son}}}\ and\ \bibinfo {author} {\bibfnamefont {A.~O.}\ \bibnamefont
  {{Starinets}}},\ }\href {\doibase 10.1146/annurev.nucl.57.090506.123120}
  {\bibfield  {journal} {\bibinfo  {journal} {Annu. Rev. Nucl. Part. S.}\
  }\textbf {\bibinfo {volume} {57}},\ \bibinfo {pages} {95} (\bibinfo {year}
  {2007})},\ \Eprint {http://arxiv.org/abs/0704.0240} {arXiv:0704.0240
  [hep-th]} \BibitemShut {NoStop}%
\bibitem [{\citenamefont {{Hartnoll}}\ \emph {et~al.}(2018)\citenamefont
  {{Hartnoll}}, \citenamefont {{Lucas}},\ and\ \citenamefont
  {{Sachdev}}}]{Hartnoll2018}%
  \BibitemOpen
  \bibfield  {author} {\bibinfo {author} {\bibfnamefont {S.~A.}\ \bibnamefont
  {{Hartnoll}}}, \bibinfo {author} {\bibfnamefont {A.}~\bibnamefont {{Lucas}}},
  \ and\ \bibinfo {author} {\bibfnamefont {S.}~\bibnamefont {{Sachdev}}},\
  }\href@noop {} {\emph {\bibinfo {title} {{Holographic quantum matter}}}}\
  (\bibinfo  {publisher} {MIT Press},\ \bibinfo {year} {2018})\BibitemShut
  {NoStop}%
\bibitem [{\citenamefont {{Hartman}}\ \emph {et~al.}(2017)\citenamefont
  {{Hartman}}, \citenamefont {{Hartnoll}},\ and\ \citenamefont
  {{Mahajan}}}]{Hartman2017}%
  \BibitemOpen
  \bibfield  {author} {\bibinfo {author} {\bibfnamefont {T.}~\bibnamefont
  {{Hartman}}}, \bibinfo {author} {\bibfnamefont {S.~A.}\ \bibnamefont
  {{Hartnoll}}}, \ and\ \bibinfo {author} {\bibfnamefont {R.}~\bibnamefont
  {{Mahajan}}},\ }\href {\doibase 10.1103/PhysRevLett.119.141601} {\bibfield
  {journal} {\bibinfo  {journal} {Phys. Rev. Lett.}\ }\textbf {\bibinfo
  {volume} {119}},\ \bibinfo {pages} {141601} (\bibinfo {year}
  {2017})}\BibitemShut {NoStop}%
\bibitem [{\citenamefont {{Hartnoll}}(2015)}]{Hartnoll2015}%
  \BibitemOpen
  \bibfield  {author} {\bibinfo {author} {\bibfnamefont {S.~A.}\ \bibnamefont
  {{Hartnoll}}},\ }\href {\doibase 10.1038/nphys3174} {\bibfield  {journal}
  {\bibinfo  {journal} {Nat. Phys.}\ }\textbf {\bibinfo {volume} {11}},\
  \bibinfo {pages} {54} (\bibinfo {year} {2015})},\ \Eprint
  {http://arxiv.org/abs/1405.3651} {arXiv:1405.3651 [cond-mat.str-el]}
  \BibitemShut {NoStop}%
\bibitem [{\citenamefont {{Gunnarsson}}\ \emph {et~al.}(2003)\citenamefont
  {{Gunnarsson}}, \citenamefont {{Calandra}},\ and\ \citenamefont
  {{Han}}}]{Gunnarsson2003}%
  \BibitemOpen
  \bibfield  {author} {\bibinfo {author} {\bibfnamefont {O.}~\bibnamefont
  {{Gunnarsson}}}, \bibinfo {author} {\bibfnamefont {M.}~\bibnamefont
  {{Calandra}}}, \ and\ \bibinfo {author} {\bibfnamefont {J.~E.}\ \bibnamefont
  {{Han}}},\ }\href {\doibase 10.1103/RevModPhys.75.1085} {\bibfield  {journal}
  {\bibinfo  {journal} {Rev. Mod. Phys.}\ }\textbf {\bibinfo {volume} {75}},\
  \bibinfo {pages} {1085} (\bibinfo {year} {2003})},\ \Eprint
  {http://arxiv.org/abs/cond-mat/0305412} {cond-mat/0305412} \BibitemShut
  {NoStop}%
\bibitem [{\citenamefont {{Bruin}}\ \emph {et~al.}(2013)\citenamefont
  {{Bruin}}, \citenamefont {{Sakai}}, \citenamefont {{Perry}},\ and\
  \citenamefont {{Mackenzie}}}]{Bruin2013}%
  \BibitemOpen
  \bibfield  {author} {\bibinfo {author} {\bibfnamefont {J.~A.~N.}\
  \bibnamefont {{Bruin}}}, \bibinfo {author} {\bibfnamefont {H.}~\bibnamefont
  {{Sakai}}}, \bibinfo {author} {\bibfnamefont {R.~S.}\ \bibnamefont
  {{Perry}}}, \ and\ \bibinfo {author} {\bibfnamefont {A.~P.}\ \bibnamefont
  {{Mackenzie}}},\ }\href {\doibase 10.1126/science.1227612} {\bibfield
  {journal} {\bibinfo  {journal} {Science}\ }\textbf {\bibinfo {volume}
  {339}},\ \bibinfo {pages} {804} (\bibinfo {year} {2013})}\BibitemShut
  {NoStop}%
\bibitem [{\citenamefont {{Damle}}\ and\ \citenamefont
  {{Sachdev}}(1997)}]{Damle1997}%
  \BibitemOpen
  \bibfield  {author} {\bibinfo {author} {\bibfnamefont {K.}~\bibnamefont
  {{Damle}}}\ and\ \bibinfo {author} {\bibfnamefont {S.}~\bibnamefont
  {{Sachdev}}},\ }\href {\doibase 10.1103/PhysRevB.56.8714} {\bibfield
  {journal} {\bibinfo  {journal} {Phys. Rev. B}\ }\textbf {\bibinfo {volume}
  {56}},\ \bibinfo {pages} {8714} (\bibinfo {year} {1997})},\ \Eprint
  {http://arxiv.org/abs/cond-mat/9705206} {cond-mat/9705206} \BibitemShut
  {NoStop}%
\bibitem [{\citenamefont {{Zaanen}}(2004)}]{Zaanen2004}%
  \BibitemOpen
  \bibfield  {author} {\bibinfo {author} {\bibfnamefont {J.}~\bibnamefont
  {{Zaanen}}},\ }\href {\doibase 10.1038/430512a} {\bibfield  {journal}
  {\bibinfo  {journal} {Nature}\ }\textbf {\bibinfo {volume} {430}},\ \bibinfo
  {pages} {512} (\bibinfo {year} {2004})}\BibitemShut {NoStop}%
\bibitem [{\citenamefont {{Lucas}}\ and\ \citenamefont
  {{Hartnoll}}(2017)}]{Lucas2017a}%
  \BibitemOpen
  \bibfield  {author} {\bibinfo {author} {\bibfnamefont {A.}~\bibnamefont
  {{Lucas}}}\ and\ \bibinfo {author} {\bibfnamefont {S.~A.}\ \bibnamefont
  {{Hartnoll}}},\ }\href {\doibase 10.1073/pnas.1711414114} {\bibfield
  {journal} {\bibinfo  {journal} {Proceedings of the National Academy of
  Science}\ }\textbf {\bibinfo {volume} {114}},\ \bibinfo {pages} {11344}
  (\bibinfo {year} {2017})},\ \Eprint {http://arxiv.org/abs/1704.07384}
  {arXiv:1704.07384 [cond-mat.str-el]} \BibitemShut {NoStop}%
\bibitem [{\citenamefont {{Lucas}}\ and\ \citenamefont
  {{Hartnoll}}(2018)}]{Lucas2018}%
  \BibitemOpen
  \bibfield  {author} {\bibinfo {author} {\bibfnamefont {A.}~\bibnamefont
  {{Lucas}}}\ and\ \bibinfo {author} {\bibfnamefont {S.~A.}\ \bibnamefont
  {{Hartnoll}}},\ }\href {\doibase 10.1103/PhysRevB.97.045105} {\bibfield
  {journal} {\bibinfo  {journal} {Phys. Rev. B}\ }\textbf {\bibinfo {volume}
  {97}},\ \bibinfo {pages} {045105} (\bibinfo {year} {2018})},\ \Eprint
  {http://arxiv.org/abs/1706.04621} {arXiv:1706.04621 [cond-mat.str-el]}
  \BibitemShut {NoStop}%
\bibitem [{\citenamefont {{Pal}}\ \emph {et~al.}(2012)\citenamefont {{Pal}},
  \citenamefont {{Yudson}},\ and\ \citenamefont {{Maslov}}}]{Pal2012}%
  \BibitemOpen
  \bibfield  {author} {\bibinfo {author} {\bibfnamefont {H.~K.}\ \bibnamefont
  {{Pal}}}, \bibinfo {author} {\bibfnamefont {V.~I.}\ \bibnamefont {{Yudson}}},
  \ and\ \bibinfo {author} {\bibfnamefont {D.~L.}\ \bibnamefont {{Maslov}}},\
  }\href {\doibase 10.3952/lithjphys.52207} {\bibfield  {journal} {\bibinfo
  {journal} {Lithuanian Journal of Physics and Technical Sciences}\ }\textbf
  {\bibinfo {volume} {52}},\ \bibinfo {pages} {142} (\bibinfo {year} {2012})},\
  \Eprint {http://arxiv.org/abs/1204.3591} {arXiv:1204.3591 [cond-mat.str-el]}
  \BibitemShut {NoStop}%
\bibitem [{\citenamefont {{Patel}}\ \emph {et~al.}(2017)\citenamefont
  {{Patel}}, \citenamefont {{Davison}},\ and\ \citenamefont
  {{Levchenko}}}]{Patel2017}%
  \BibitemOpen
  \bibfield  {author} {\bibinfo {author} {\bibfnamefont {A.~A.}\ \bibnamefont
  {{Patel}}}, \bibinfo {author} {\bibfnamefont {R.~A.}\ \bibnamefont
  {{Davison}}}, \ and\ \bibinfo {author} {\bibfnamefont {A.}~\bibnamefont
  {{Levchenko}}},\ }\href {\doibase 10.1103/PhysRevB.96.205417} {\bibfield
  {journal} {\bibinfo  {journal} {Phys. Rev. B}\ }\textbf {\bibinfo {volume}
  {96}},\ \bibinfo {pages} {205417} (\bibinfo {year} {2017})},\ \Eprint
  {http://arxiv.org/abs/1706.03775} {arXiv:1706.03775 [cond-mat.str-el]}
  \BibitemShut {NoStop}%
\bibitem [{\citenamefont {{Werman}}\ and\ \citenamefont
  {{Berg}}(2016)}]{Werman2016}%
  \BibitemOpen
  \bibfield  {author} {\bibinfo {author} {\bibfnamefont {Y.}~\bibnamefont
  {{Werman}}}\ and\ \bibinfo {author} {\bibfnamefont {E.}~\bibnamefont
  {{Berg}}},\ }\href {\doibase 10.1103/PhysRevB.93.075109} {\bibfield
  {journal} {\bibinfo  {journal} {Phys. Rev. B}\ }\textbf {\bibinfo {volume}
  {93}},\ \bibinfo {pages} {075109} (\bibinfo {year} {2016})},\ \Eprint
  {http://arxiv.org/abs/1512.00041} {arXiv:1512.00041 [cond-mat.mes-hall]}
  \BibitemShut {NoStop}%
\bibitem [{\citenamefont {{Werman}}\ \emph {et~al.}(2017)\citenamefont
  {{Werman}}, \citenamefont {{Kivelson}},\ and\ \citenamefont
  {{Berg}}}]{Werman2017}%
  \BibitemOpen
  \bibfield  {author} {\bibinfo {author} {\bibfnamefont {Y.}~\bibnamefont
  {{Werman}}}, \bibinfo {author} {\bibfnamefont {S.~A.}\ \bibnamefont
  {{Kivelson}}}, \ and\ \bibinfo {author} {\bibfnamefont {E.}~\bibnamefont
  {{Berg}}},\ }\href {\doibase 10.1038/s41535-017-0009-8} {\bibfield  {journal}
  {\bibinfo  {journal} {npj Quant. Mats.}\ }\textbf {\bibinfo {volume} {2}},\
  \bibinfo {pages} {7} (\bibinfo {year} {2017})},\ \Eprint
  {http://arxiv.org/abs/1607.05725} {arXiv:1607.05725 [cond-mat.str-el]}
  \BibitemShut {NoStop}%
\bibitem [{\citenamefont {{Song}}\ \emph {et~al.}(2017)\citenamefont {{Song}},
  \citenamefont {{Jian}},\ and\ \citenamefont {{Balents}}}]{Song2017}%
  \BibitemOpen
  \bibfield  {author} {\bibinfo {author} {\bibfnamefont {X.-Y.}\ \bibnamefont
  {{Song}}}, \bibinfo {author} {\bibfnamefont {C.-M.}\ \bibnamefont {{Jian}}},
  \ and\ \bibinfo {author} {\bibfnamefont {L.}~\bibnamefont {{Balents}}},\
  }\href {\doibase 10.1103/PhysRevLett.119.216601} {\bibfield  {journal}
  {\bibinfo  {journal} {Phys. Rev. Lett.}\ }\textbf {\bibinfo {volume} {119}},\
  \bibinfo {pages} {216601} (\bibinfo {year} {2017})},\ \Eprint
  {http://arxiv.org/abs/1705.00117} {arXiv:1705.00117 [cond-mat.str-el]}
  \BibitemShut {NoStop}%
\bibitem [{\citenamefont {{Chowdhury}}\ \emph {et~al.}(2018)\citenamefont
  {{Chowdhury}}, \citenamefont {{Werman}}, \citenamefont {{Berg}},\ and\
  \citenamefont {{Senthil}}}]{Chowdhury2018}%
  \BibitemOpen
  \bibfield  {author} {\bibinfo {author} {\bibfnamefont {D.}~\bibnamefont
  {{Chowdhury}}}, \bibinfo {author} {\bibfnamefont {Y.}~\bibnamefont
  {{Werman}}}, \bibinfo {author} {\bibfnamefont {E.}~\bibnamefont {{Berg}}}, \
  and\ \bibinfo {author} {\bibfnamefont {T.}~\bibnamefont {{Senthil}}},\
  }\href@noop {} {\bibfield  {journal} {\bibinfo  {journal} {arXiv}\ }
  (\bibinfo {year} {2018})},\ \Eprint {http://arxiv.org/abs/1801.06178}
  {arXiv:1801.06178 [cond-mat.str-el]} \BibitemShut {NoStop}%
\bibitem [{\citenamefont {{Lucas}}(2017)}]{Lucas2017}%
  \BibitemOpen
  \bibfield  {author} {\bibinfo {author} {\bibfnamefont {A.}~\bibnamefont
  {{Lucas}}},\ }\href@noop {} {\bibfield  {journal} {\bibinfo  {journal}
  {arXiv}\ ,\ \bibinfo {pages} {1710.01005}} (\bibinfo {year} {2017})},\
  \Eprint {http://arxiv.org/abs/1710.01005} {arXiv:1710.01005 [hep-th]}
  \BibitemShut {NoStop}%
\bibitem [{\citenamefont {{Apostolov}}\ \emph {et~al.}(2014)\citenamefont
  {{Apostolov}}, \citenamefont {{Levchenko}},\ and\ \citenamefont
  {{Andreev}}}]{Apostolov2014}%
  \BibitemOpen
  \bibfield  {author} {\bibinfo {author} {\bibfnamefont {S.~S.}\ \bibnamefont
  {{Apostolov}}}, \bibinfo {author} {\bibfnamefont {A.}~\bibnamefont
  {{Levchenko}}}, \ and\ \bibinfo {author} {\bibfnamefont {A.~V.}\ \bibnamefont
  {{Andreev}}},\ }\href {\doibase 10.1103/PhysRevB.89.121104} {\bibfield
  {journal} {\bibinfo  {journal} {Phys. Rev. B}\ }\textbf {\bibinfo {volume}
  {89}},\ \bibinfo {pages} {121104} (\bibinfo {year} {2014})},\ \Eprint
  {http://arxiv.org/abs/1312.6890} {arXiv:1312.6890 [cond-mat.str-el]}
  \BibitemShut {NoStop}%
\bibitem [{\citenamefont {{Chen}}\ \emph {et~al.}(2015)\citenamefont {{Chen}},
  \citenamefont {{Andreev}},\ and\ \citenamefont {{Levchenko}}}]{Chen2015}%
  \BibitemOpen
  \bibfield  {author} {\bibinfo {author} {\bibfnamefont {W.}~\bibnamefont
  {{Chen}}}, \bibinfo {author} {\bibfnamefont {A.~V.}\ \bibnamefont
  {{Andreev}}}, \ and\ \bibinfo {author} {\bibfnamefont {A.}~\bibnamefont
  {{Levchenko}}},\ }\href {\doibase 10.1103/PhysRevB.91.245405} {\bibfield
  {journal} {\bibinfo  {journal} {Phys. Rev. B}\ }\textbf {\bibinfo {volume}
  {91}},\ \bibinfo {pages} {245405} (\bibinfo {year} {2015})},\ \Eprint
  {http://arxiv.org/abs/1503.05566} {arXiv:1503.05566 [cond-mat.mes-hall]}
  \BibitemShut {NoStop}%
\bibitem [{\citenamefont {{Spivak}}\ and\ \citenamefont
  {{Kivelson}}(2006)}]{Spivak2006}%
  \BibitemOpen
  \bibfield  {author} {\bibinfo {author} {\bibfnamefont {B.}~\bibnamefont
  {{Spivak}}}\ and\ \bibinfo {author} {\bibfnamefont {S.~A.}\ \bibnamefont
  {{Kivelson}}},\ }\href {\doibase 10.1016/j.aop.2005.12.002} {\bibfield
  {journal} {\bibinfo  {journal} {Ann. Phys.}\ }\textbf {\bibinfo {volume}
  {321}},\ \bibinfo {pages} {2071} (\bibinfo {year} {2006})}\BibitemShut
  {NoStop}%
\bibitem [{\citenamefont {{Andreev}}\ \emph {et~al.}(2011)\citenamefont
  {{Andreev}}, \citenamefont {{Kivelson}},\ and\ \citenamefont
  {{Spivak}}}]{Andreev2011}%
  \BibitemOpen
  \bibfield  {author} {\bibinfo {author} {\bibfnamefont {A.~V.}\ \bibnamefont
  {{Andreev}}}, \bibinfo {author} {\bibfnamefont {S.~A.}\ \bibnamefont
  {{Kivelson}}}, \ and\ \bibinfo {author} {\bibfnamefont {B.}~\bibnamefont
  {{Spivak}}},\ }\href {\doibase 10.1103/PhysRevLett.106.256804} {\bibfield
  {journal} {\bibinfo  {journal} {Phys. Rev. Lett.}\ }\textbf {\bibinfo
  {volume} {106}},\ \bibinfo {pages} {256804} (\bibinfo {year} {2011})},\
  \Eprint {http://arxiv.org/abs/1011.3068} {arXiv:1011.3068
  [cond-mat.mes-hall]} \BibitemShut {NoStop}%
\bibitem [{\citenamefont {Ziman}(1960)}]{Ziman1960}%
  \BibitemOpen
  \bibfield  {author} {\bibinfo {author} {\bibfnamefont {J.~M.}\ \bibnamefont
  {Ziman}},\ }\href@noop {} {\emph {\bibinfo {title} {Electrons and Phonons}}}\
  (\bibinfo  {publisher} {Clarendon Press},\ \bibinfo {year}
  {1960})\BibitemShut {NoStop}%
\bibitem [{\citenamefont {{Mousatov}}\ \emph {et~al.}(2018)\citenamefont
  {{Mousatov}}, \citenamefont {{Esterlis}},\ and\ \citenamefont
  {{Hartnoll}}}]{Mousatov2018}%
  \BibitemOpen
  \bibfield  {author} {\bibinfo {author} {\bibfnamefont {C.~H.}\ \bibnamefont
  {{Mousatov}}}, \bibinfo {author} {\bibfnamefont {I.}~\bibnamefont
  {{Esterlis}}}, \ and\ \bibinfo {author} {\bibfnamefont {S.~A.}\ \bibnamefont
  {{Hartnoll}}},\ }\href@noop {} {\bibfield  {journal} {\bibinfo  {journal}
  {arXiv}\ } (\bibinfo {year} {2018})},\ \Eprint
  {http://arxiv.org/abs/1803.08054} {arXiv:1803.08054 [cond-mat.str-el]}
  \BibitemShut {NoStop}%
\bibitem [{\citenamefont {{Andreev}}(1978)}]{Andreev1978}%
  \BibitemOpen
  \bibfield  {author} {\bibinfo {author} {\bibfnamefont {A.~F.}\ \bibnamefont
  {{Andreev}}},\ }\href@noop {} {\bibfield  {journal} {\bibinfo  {journal}
  {Soviet Journal of Experimental and Theoretical Physics Letters}\ }\textbf
  {\bibinfo {volume} {28}},\ \bibinfo {pages} {556} (\bibinfo {year}
  {1978})}\BibitemShut {NoStop}%
\bibitem [{\citenamefont {{Andreev}}\ and\ \citenamefont
  {{Kosevich}}(1979)}]{Andreev1979}%
  \BibitemOpen
  \bibfield  {author} {\bibinfo {author} {\bibfnamefont {A.~F.}\ \bibnamefont
  {{Andreev}}}\ and\ \bibinfo {author} {\bibfnamefont {Y.~A.}\ \bibnamefont
  {{Kosevich}}},\ }\href@noop {} {\bibfield  {journal} {\bibinfo  {journal}
  {Soviet Journal of Experimental and Theoretical Physics}\ }\textbf {\bibinfo
  {volume} {50}},\ \bibinfo {pages} {1218} (\bibinfo {year}
  {1979})}\BibitemShut {NoStop}%
\bibitem [{\citenamefont {{Narozhny}}\ and\ \citenamefont
  {{Levchenko}}(2016)}]{Narozhny2016}%
  \BibitemOpen
  \bibfield  {author} {\bibinfo {author} {\bibfnamefont {B.~N.}\ \bibnamefont
  {{Narozhny}}}\ and\ \bibinfo {author} {\bibfnamefont {A.}~\bibnamefont
  {{Levchenko}}},\ }\href {\doibase 10.1103/RevModPhys.88.025003} {\bibfield
  {journal} {\bibinfo  {journal} {Rev. Mod. Phys.}\ }\textbf {\bibinfo {volume}
  {88}},\ \bibinfo {pages} {025003} (\bibinfo {year} {2016})},\ \Eprint
  {http://arxiv.org/abs/1505.07468} {arXiv:1505.07468 [cond-mat.mes-hall]}
  \BibitemShut {NoStop}%
\bibitem [{\citenamefont {{Braude}}\ and\ \citenamefont
  {{Stern}}(2001)}]{Braude2001}%
  \BibitemOpen
  \bibfield  {author} {\bibinfo {author} {\bibfnamefont {V.}~\bibnamefont
  {{Braude}}}\ and\ \bibinfo {author} {\bibfnamefont {A.}~\bibnamefont
  {{Stern}}},\ }\href {\doibase 10.1103/PhysRevB.64.115431} {\bibfield
  {journal} {\bibinfo  {journal} {Phys. Rev. B}\ }\textbf {\bibinfo {volume}
  {64}},\ \bibinfo {pages} {115431} (\bibinfo {year} {2001})},\ \Eprint
  {http://arxiv.org/abs/cond-mat/0103306} {cond-mat/0103306} \BibitemShut
  {NoStop}%
\bibitem [{\citenamefont {{Kamenev}}\ and\ \citenamefont
  {{Oreg}}(1995)}]{Kamenev1995}%
  \BibitemOpen
  \bibfield  {author} {\bibinfo {author} {\bibfnamefont {A.}~\bibnamefont
  {{Kamenev}}}\ and\ \bibinfo {author} {\bibfnamefont {Y.}~\bibnamefont
  {{Oreg}}},\ }\href {\doibase 10.1103/PhysRevB.52.7516} {\bibfield  {journal}
  {\bibinfo  {journal} {Phys. Rev. B}\ }\textbf {\bibinfo {volume} {52}},\
  \bibinfo {pages} {7516} (\bibinfo {year} {1995})},\ \Eprint
  {http://arxiv.org/abs/cond-mat/9504057} {cond-mat/9504057} \BibitemShut
  {NoStop}%
\bibitem [{\citenamefont {{Flensberg}}\ \emph {et~al.}(1995)\citenamefont
  {{Flensberg}}, \citenamefont {{Hu}}, \citenamefont {{Jauho}},\ and\
  \citenamefont {{Kinaret}}}]{Flensberg1995}%
  \BibitemOpen
  \bibfield  {author} {\bibinfo {author} {\bibfnamefont {K.}~\bibnamefont
  {{Flensberg}}}, \bibinfo {author} {\bibfnamefont {B.~Y.-K.}\ \bibnamefont
  {{Hu}}}, \bibinfo {author} {\bibfnamefont {A.-P.}\ \bibnamefont {{Jauho}}}, \
  and\ \bibinfo {author} {\bibfnamefont {J.~M.}\ \bibnamefont {{Kinaret}}},\
  }\href {\doibase 10.1103/PhysRevB.52.14761} {\bibfield  {journal} {\bibinfo
  {journal} {Phys. Rev. B}\ }\textbf {\bibinfo {volume} {52}},\ \bibinfo
  {pages} {14761} (\bibinfo {year} {1995})},\ \Eprint
  {http://arxiv.org/abs/cond-mat/9504092} {cond-mat/9504092} \BibitemShut
  {NoStop}%
\bibitem [{\citenamefont {Friedrich}(2013)}]{Friedrich2013}%
  \BibitemOpen
  \bibfield  {author} {\bibinfo {author} {\bibfnamefont {H.}~\bibnamefont
  {Friedrich}},\ }\href@noop {} {\emph {\bibinfo {title} {Scattering Theory}}}\
  (\bibinfo  {publisher} {Springer},\ \bibinfo {year} {2013})\BibitemShut
  {NoStop}%
\bibitem [{\citenamefont {{Lucas}}\ and\ \citenamefont {{Das
  Sarma}}(2018)}]{Lucas2018a}%
  \BibitemOpen
  \bibfield  {author} {\bibinfo {author} {\bibfnamefont {A.}~\bibnamefont
  {{Lucas}}}\ and\ \bibinfo {author} {\bibfnamefont {S.}~\bibnamefont {{Das
  Sarma}}},\ }\href {\doibase 10.1103/PhysRevB.97.115449} {\bibfield  {journal}
  {\bibinfo  {journal} {Phys. Rev. B}\ }\textbf {\bibinfo {volume} {97}},\
  \bibinfo {pages} {115449} (\bibinfo {year} {2018})},\ \Eprint
  {http://arxiv.org/abs/1801.01495} {arXiv:1801.01495 [cond-mat.str-el]}
  \BibitemShut {NoStop}%
\bibitem [{\citenamefont {Ioffe}\ and\ \citenamefont
  {Regel}(1960)}]{Ioffe1960}%
  \BibitemOpen
  \bibfield  {author} {\bibinfo {author} {\bibfnamefont {A.~F.}\ \bibnamefont
  {Ioffe}}\ and\ \bibinfo {author} {\bibfnamefont {A.~R.}\ \bibnamefont
  {Regel}},\ }\href@noop {} {\bibfield  {journal} {\bibinfo  {journal}
  {Prog.Semicond., Vol.4}\ ,\ \bibinfo {pages} {237}} (\bibinfo {year}
  {1960})}\BibitemShut {NoStop}%
\bibitem [{\citenamefont {{Mahajan}}\ \emph {et~al.}(2013)\citenamefont
  {{Mahajan}}, \citenamefont {{Barkeshli}},\ and\ \citenamefont
  {{Hartnoll}}}]{Mahajan2013a}%
  \BibitemOpen
  \bibfield  {author} {\bibinfo {author} {\bibfnamefont {R.}~\bibnamefont
  {{Mahajan}}}, \bibinfo {author} {\bibfnamefont {M.}~\bibnamefont
  {{Barkeshli}}}, \ and\ \bibinfo {author} {\bibfnamefont {S.~A.}\ \bibnamefont
  {{Hartnoll}}},\ }\href {\doibase 10.1103/PhysRevB.88.125107} {\bibfield
  {journal} {\bibinfo  {journal} {Phys. Rev. B}\ }\textbf {\bibinfo {volume}
  {88}},\ \bibinfo {pages} {125107} (\bibinfo {year} {2013})},\ \Eprint
  {http://arxiv.org/abs/1304.4249} {arXiv:1304.4249 [cond-mat.str-el]}
  \BibitemShut {NoStop}%
\bibitem [{\citenamefont {{Jaoui}}\ \emph {et~al.}(2018)\citenamefont
  {{Jaoui}}, \citenamefont {{Fauqu{\'e}}}, \citenamefont {{Rischau}},
  \citenamefont {{Subedi}}, \citenamefont {{Fu}}, \citenamefont {{Gooth}},
  \citenamefont {{Kumar}}, \citenamefont {{S{\"u}{\ss}}}, \citenamefont
  {{Maslov}}, \citenamefont {{Felser}},\ and\ \citenamefont
  {{Behnia}}}]{Jaoui2018}%
  \BibitemOpen
  \bibfield  {author} {\bibinfo {author} {\bibfnamefont {A.}~\bibnamefont
  {{Jaoui}}}, \bibinfo {author} {\bibfnamefont {B.}~\bibnamefont
  {{Fauqu{\'e}}}}, \bibinfo {author} {\bibfnamefont {C.~W.}\ \bibnamefont
  {{Rischau}}}, \bibinfo {author} {\bibfnamefont {A.}~\bibnamefont {{Subedi}}},
  \bibinfo {author} {\bibfnamefont {C.}~\bibnamefont {{Fu}}}, \bibinfo {author}
  {\bibfnamefont {J.}~\bibnamefont {{Gooth}}}, \bibinfo {author} {\bibfnamefont
  {N.}~\bibnamefont {{Kumar}}}, \bibinfo {author} {\bibfnamefont
  {V.}~\bibnamefont {{S{\"u}{\ss}}}}, \bibinfo {author} {\bibfnamefont {D.~L.}\
  \bibnamefont {{Maslov}}}, \bibinfo {author} {\bibfnamefont {C.}~\bibnamefont
  {{Felser}}}, \ and\ \bibinfo {author} {\bibfnamefont {K.}~\bibnamefont
  {{Behnia}}},\ }\href {\doibase 10.1038/s41535-018-0136-x} {\bibfield
  {journal} {\bibinfo  {journal} {npj Quant. Mats.}\ }\textbf {\bibinfo
  {volume} {3}},\ \bibinfo {pages} {64} (\bibinfo {year} {2018})},\ \Eprint
  {http://arxiv.org/abs/1806.04094} {arXiv:1806.04094 [cond-mat.str-el]}
  \BibitemShut {NoStop}%
\bibitem [{\citenamefont {{Gooth}}\ \emph {et~al.}(2018)\citenamefont
  {{Gooth}}, \citenamefont {{Menges}}, \citenamefont {{Kumar}}, \citenamefont
  {{S{\"u}{\ss}}}, \citenamefont {{Shekhar}}, \citenamefont {{Sun}},
  \citenamefont {{Drechsler}}, \citenamefont {{Zierold}}, \citenamefont
  {{Felser}},\ and\ \citenamefont {{Gotsmann}}}]{Gooth2017}%
  \BibitemOpen
  \bibfield  {author} {\bibinfo {author} {\bibfnamefont {J.}~\bibnamefont
  {{Gooth}}}, \bibinfo {author} {\bibfnamefont {F.}~\bibnamefont {{Menges}}},
  \bibinfo {author} {\bibfnamefont {N.}~\bibnamefont {{Kumar}}}, \bibinfo
  {author} {\bibfnamefont {V.}~\bibnamefont {{S{\"u}{\ss}}}}, \bibinfo {author}
  {\bibfnamefont {C.}~\bibnamefont {{Shekhar}}}, \bibinfo {author}
  {\bibfnamefont {Y.}~\bibnamefont {{Sun}}}, \bibinfo {author} {\bibfnamefont
  {U.}~\bibnamefont {{Drechsler}}}, \bibinfo {author} {\bibfnamefont
  {R.}~\bibnamefont {{Zierold}}}, \bibinfo {author} {\bibfnamefont
  {C.}~\bibnamefont {{Felser}}}, \ and\ \bibinfo {author} {\bibfnamefont
  {B.}~\bibnamefont {{Gotsmann}}},\ }\href {\doibase
  10.1038/s41467-018-06688-y} {\bibfield  {journal} {\bibinfo  {journal} {Nat.
  Comm.}\ }\textbf {\bibinfo {volume} {9}},\ \bibinfo {pages} {4093} (\bibinfo
  {year} {2018})},\ \Eprint {http://arxiv.org/abs/1706.05925}
  {arXiv:1706.05925} \BibitemShut {NoStop}%
\bibitem [{\citenamefont {{Bard}}\ \emph {et~al.}(2018)\citenamefont {{Bard}},
  \citenamefont {{Protopopov}},\ and\ \citenamefont {{Mirlin}}}]{Bard2018}%
  \BibitemOpen
  \bibfield  {author} {\bibinfo {author} {\bibfnamefont {M.}~\bibnamefont
  {{Bard}}}, \bibinfo {author} {\bibfnamefont {I.~V.}\ \bibnamefont
  {{Protopopov}}}, \ and\ \bibinfo {author} {\bibfnamefont {A.~D.}\
  \bibnamefont {{Mirlin}}},\ }\href {\doibase 10.1103/PhysRevB.97.195147}
  {\bibfield  {journal} {\bibinfo  {journal} {Phys. Rev. B}\ }\textbf {\bibinfo
  {volume} {97}},\ \bibinfo {pages} {195147} (\bibinfo {year} {2018})},\
  \Eprint {http://arxiv.org/abs/1802.10429} {arXiv:1802.10429
  [cond-mat.mes-hall]} \BibitemShut {NoStop}%
\bibitem [{\citenamefont {{Barton}}(1983)}]{Barton1983}%
  \BibitemOpen
  \bibfield  {author} {\bibinfo {author} {\bibfnamefont {G.}~\bibnamefont
  {{Barton}}},\ }\href {\doibase 10.1119/1.13228} {\bibfield  {journal}
  {\bibinfo  {journal} {Am. J. Phys}\ }\textbf {\bibinfo {volume} {51}},\
  \bibinfo {pages} {420} (\bibinfo {year} {1983})}\BibitemShut {NoStop}%
\bibitem [{\citenamefont {{Goldston}}\ and\ \citenamefont
  {{Rutherford}}(1995)}]{Goldston1995}%
  \BibitemOpen
  \bibfield  {author} {\bibinfo {author} {\bibfnamefont {R.~J.}\ \bibnamefont
  {{Goldston}}}\ and\ \bibinfo {author} {\bibfnamefont {P.~H.}\ \bibnamefont
  {{Rutherford}}},\ }\href@noop {} {\emph {\bibinfo {title} {Introduction to
  Plasma Physics}}}\ (\bibinfo  {publisher} {IOP Publishing},\ \bibinfo {year}
  {1995})\BibitemShut {NoStop}%
\end{thebibliography}
\end{document}